\documentclass[final,5p,times,twocolumn]{elsarticle}
\pdfoutput=1
\usepackage{graphics}
\usepackage{amssymb}

\begin{document}

\begin{frontmatter}

\title{Dark matter searches with IceCube}

\author{Carlos de los Heros\\ for the IceCube Collaboration}

\address{Department of Physics and Astronomy. Uppsala University. Sweden}

\begin{abstract}
The construction of the IceCube neutrino observatory is practically terminated. At the time of this 
writing, and with 79 strings taking data out of the 86 foreseen, we are one deployment season away from  
completion. The detector, however, has been taking data since 2006 in different partial 
configurations. We have evaluated these data for evidence of dark matter annihilations in 
the Sun, in the Galactic Center and in the Galactic Halo, searching for an excess neutrino 
flux over the expected atmospheric neutrino background. 
   This contribution reviews the results of dark matter searches for WIMPs, Kaluza-Klein modes 
and superheavy candidates (Simpzillas), using the 22- and 40-string configurations of IceCube. The results 
are presented in the form of muon flux limits, constrains on the candidates' spin-dependent  
cross-section with protons, and constrains in the self-annihilation cross section. These results are 
presented in the context of direct searches and searches in space.
\end{abstract}

\end{frontmatter}

\section{Dark matter searches with IceCube}
 The exciting possibility of detecting dark matter particle candidates with IceCube is based on the assumption 
that, if they constitute the dark matter in the halo, they can be gravitationally trapped in the 
deep gravitational wells of heavy objects, like the Sun or the Galactic Center~\cite{DMSun}. Subsequent pair-wise annihilation into 
Standard Model particles could lead to a detectable neutrino flux. This is a clear signal for a neutrino telescope: 
it is directional and has a different energy spectrum than the atmospheric neutrino background flux. 
 Popular dark matter candidates are stable relic particles that arise in supersymmetric extensions of the Standard Model 
or in theories with extra spacial dimensions. In some flavours of the minimal supersymmetric extension of the Standard Model, 
the MSSM, a viable dark matter candidate is the lightest neutralino, the lightest particle in the 
super-partner quadruplet of the gauge bosons and neutral Higgs. Neutralinos are stable, interact only weakly and 
gravitationally and, as relics of the Big Bang, can form a dark matter halo in the galaxy. 
The current lower limit on the neutralino mass, $m_{\chi}\gtrsim 46$ GeV, comes from accelerator searches~\cite{DELPHI:03a}, 
while an upper limit of a few hundred TeV can be set based on unitarity constrains on the mass of any thermally produced 
relic~\cite{Griest:90a}. We will not discuss further other common supersymmetric scenarios where the gravitino 
is the lightest supersymmetric particle, since they do not provide a signal in neutrino 
 telescopes, but we will consider another thermal relic arising in the scenario of universal extra dimensions, 
the lightest Kaluza-Klein particle (LKP)~\cite{Hooper:07a}. We have considered the LKP in models with one additional 
space dimension, associated with the first excitation of the hypercharge gauge boson. The mass of the LKP is 
inversely proportional to the 'size' of the extra dimension and can lie in the range few hundred GeV to about a TeV. 
The model thus defined has only two parameters; the LKP mass and  the relative mass 
difference, $\Delta_q$, between the LKP and the first Kaluza-Klein quark excitation. This parameter controls the strength 
of possible co annihilations and influences the predicted relic density of LKPs. 
A third kind of candidates we have considered are Simpzillas, superheavy dark matter in the form of strongly-interacting 
relic particles in the mass range 10$^4$~GeV -- 10$^{18}$~GeV. Strongly-interacting in this context simply means non-weakly  
(as opposed to the usual assumption for WIMPs) and it should not be understood as a QCD-like interaction. Unlike neutralinos 
or LKPs, Simpzillas are produced  non-thermally at the end of inflation~\cite{Chung:98a}, and the unitarity constraint on 
their mass can therefore be avoided. Masses up to the unification scale can be generated without violating any fundamental law.

\section{The IceCube detector}
IceCube detects Cherenkov light from relativistic particles produced in neutrino interactions in or near the 
detector, using a three-dimensional array of light sensors. The nominal IceCube configuration consists of 80 vertical strings 
with 60 Digital Optical Modules (DOM) each, deployed between 1450~m and 2450~m under the ice in the South Pole glacier. 
The typical inter-string 
separation is 125~m and the DOMs are vertically spaced by 17~m within each string. Each DOM consists of a 23~cm Hamamatsu 
photomultiplier tube with digitizing electronics and a flasher board for calibration purposes, enclosed in a glass 
pressure sphere~\cite{IceCube_DOM}. Each DOM functions as an autonomous data collection unit with a 
time resolution of $\lesssim$ 2~ns.
Such geometry is optimized to detect high-energy neutrinos ($E_{\nu}\gtrsim TeV$) from cosmic ray sources. \par
To this original configuration, six additional closely-spaced strings were recently deployed 
in order to lower the energy threshold of the detector to few 10s of GeV, extending consequently its science 
potential~\cite{DeepCore}. These additional strings are equipped with  higher quantum efficiency photomultiplier 
tubes from Hamamatsu and were deployed as to form 
a denser core deep in the center of the IceCube array, with a typical inter-string separation of 72~m. 
The DOMs in the additional strings have vertical spacings between 10~m and 7~m, depending on their  depth. This 
distribution is intended to avoid instrumenting a horizontal ice layer of worse optical properties due to a 
higher content of dust, which lies approximately at a depth between 1950~m and 2080~m. 
The upper layer of DOMs, along with DOMs in the surrounding IceCube strings, can be used to define a veto region  
for contained or starting tracks. This allows to significantly reduce the energy threshold of the detector and 
to perform searches on the full sky, increasing the field of view of IceCube to the southern hemisphere and to 
the Galactic Center, as well as to be able to monitor the Sun for dark matter searches year around.\par

At the time of this conference, IceCube is taking data with 79 strings, but here we discuss 
the results obtained with the data taken with the 22-string detector in 2007 and with the 40-string detector  
between 2008 and early 2009.  The surface layout of these configurations is shown in figure~\ref{fig:IC40geo}. 
 
\begin{figure}[t]
\centering\includegraphics[width=\linewidth,height=0.9\linewidth]{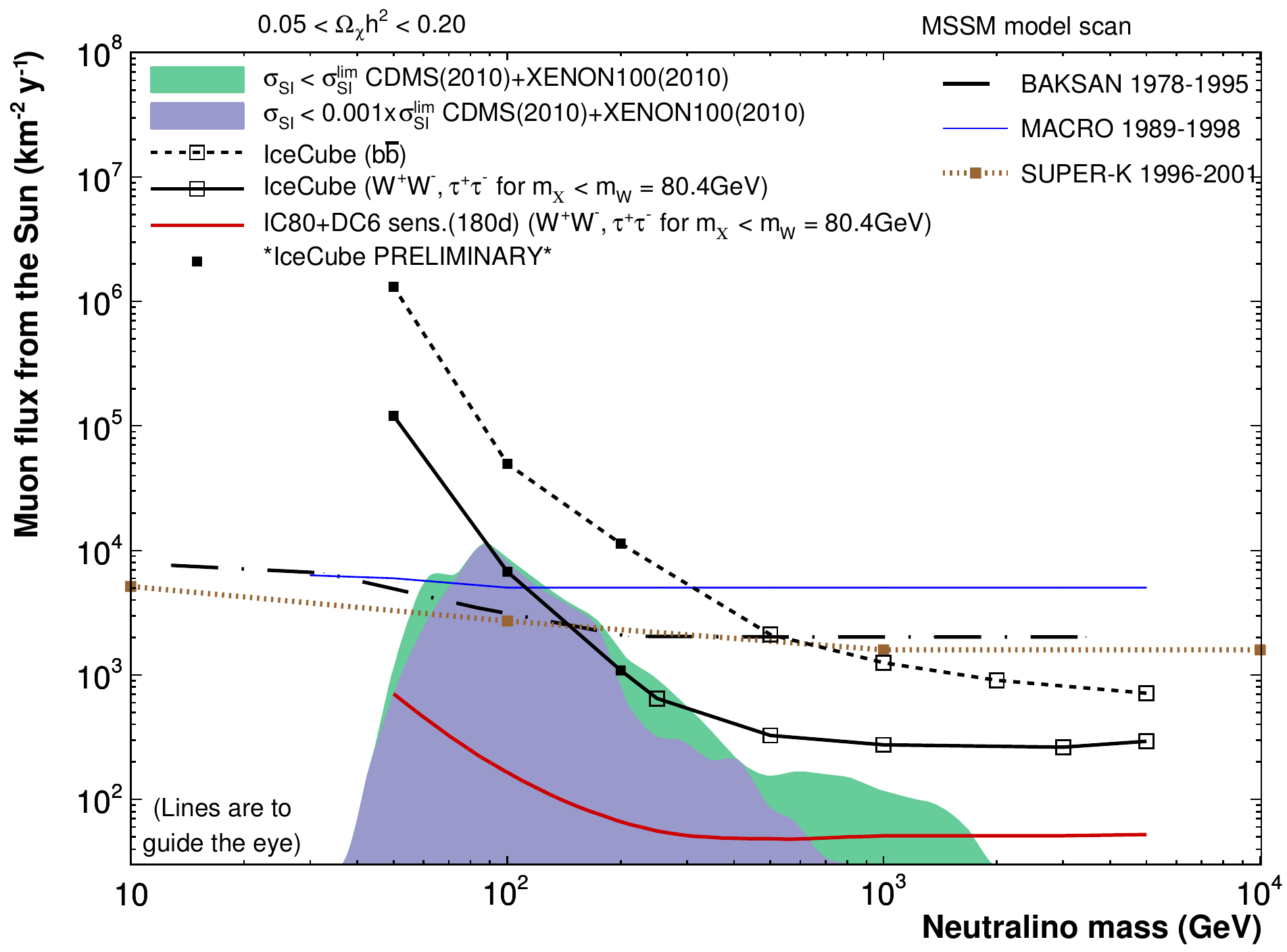}
\caption{90\% CL limits on the muon flux from neutralino annihilations in the Sun obtained with the 22-string IceCube detector (open squares), compared with 
previous results of SuperK, Baksan and MACRO~\cite{SuperK_WIMP, Baksan_WIMP, MACRO_WIMP} and the expected sensitivity 
of IceCube with the six additional DeepCore strings. The low mass results (black squares) were obtained with AMANDA~\cite{AMANDA_WIMP} and are 
preliminary.}
\label{fig:IC22_flux_limits}
\end{figure}

\section{Searches for dark matter signals from the Sun}
In this section we summarize the results obtained with the 22-string configuration of IceCube on the search for MSSM 
neutralino, LKP and Simpzilla dark matter. The data set used was collected between March and September 2007, when the 
Sun was below the horizon, and consists of 4.8$\times$10$^9$ events at trigger level, recorded in 104~days of detector live time.  
After a series of quality cuts to reduce the background, mainly consisting of misreconstructed 
 atmospheric muons, 6946 neutrino candidates remain. This is compatible with the number of events expected from the known atmospheric 
neutrino flux. A statistical test based on the angular distribution of the data with respect to the direction of the Sun was used to 
extract muon flux limits from the annihilations of each of the dark matter candidates considered. Systematic uncertainties in this 
analysis lie in the range 19\%-26\%, and arise mainly from uncertainties in the total efficiency of the optical modules and in the 
simulation of the optical properties of the ice. They have been included in the calculation of the limits presented below.
\begin{figure}[t]
\centering\includegraphics[width=\linewidth,height=0.9\linewidth]{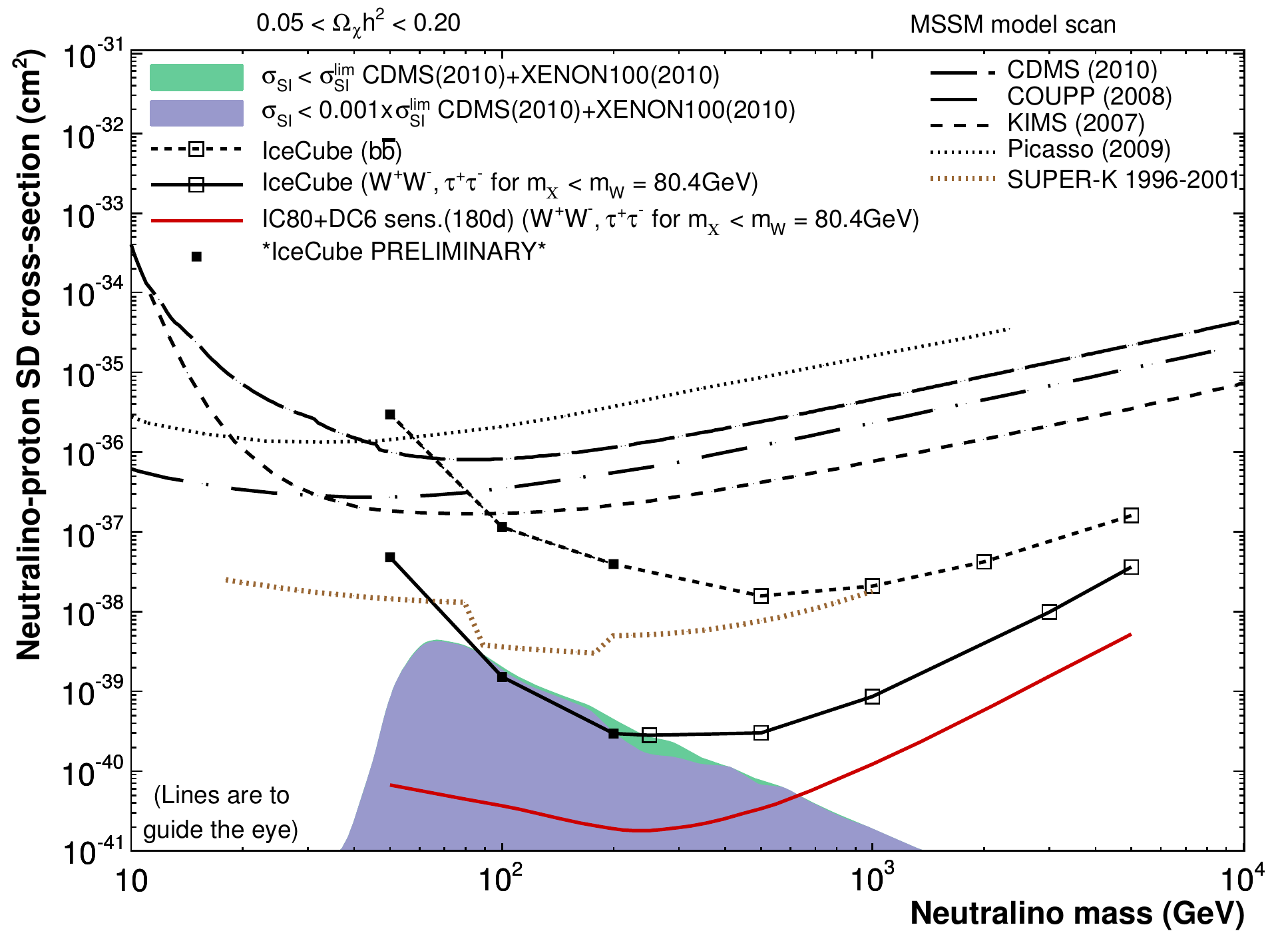}
\caption{90\% CL limits on the spin-dependent WIMP-proton cross section obtained with 22-string IceCube detector, compared with direct search 
experiments~\cite{CDMS,COUPP,KIMS} and Super-K~\cite{SuperK_WIMP}. The low mass results (black squares) were obtained with AMANDA~\cite{AMANDA_WIMP} and are 
preliminary.}
\label{fig:IC22_Xsection_limits}
\end{figure}

\begin{figure}[t]
\centering\includegraphics[width=\linewidth,height=0.9\linewidth]{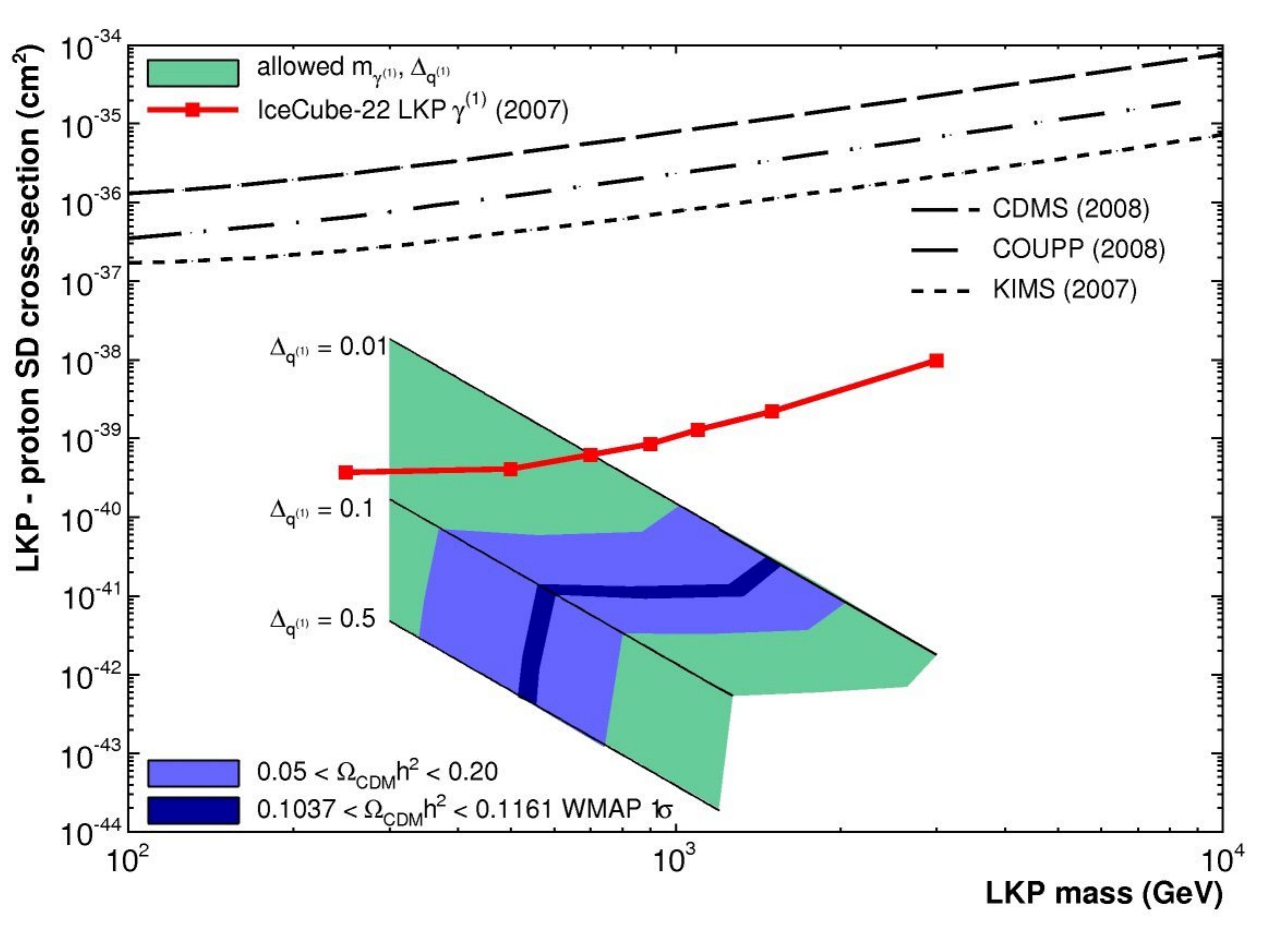}
\caption{90\% CL limits on the spin-dependent LKP-proton cross section obtained with 22-string IceCube detector. The theoretically allowed 
parameter space as a function of two choice values of $\Delta_q$ is indicated as the green area. The blue regions 
correspond to two values of allowed dark matter relic density according to the WMAP results. Limits from direct experiments~\cite{CDMS,COUPP,KIMS} 
on WIMP-proton cross section are shown for comparison.}
\label{fig:LKP_limit}
\end{figure}

\begin{figure}[t]
\centering\includegraphics[width=\linewidth,height=0.9\linewidth]{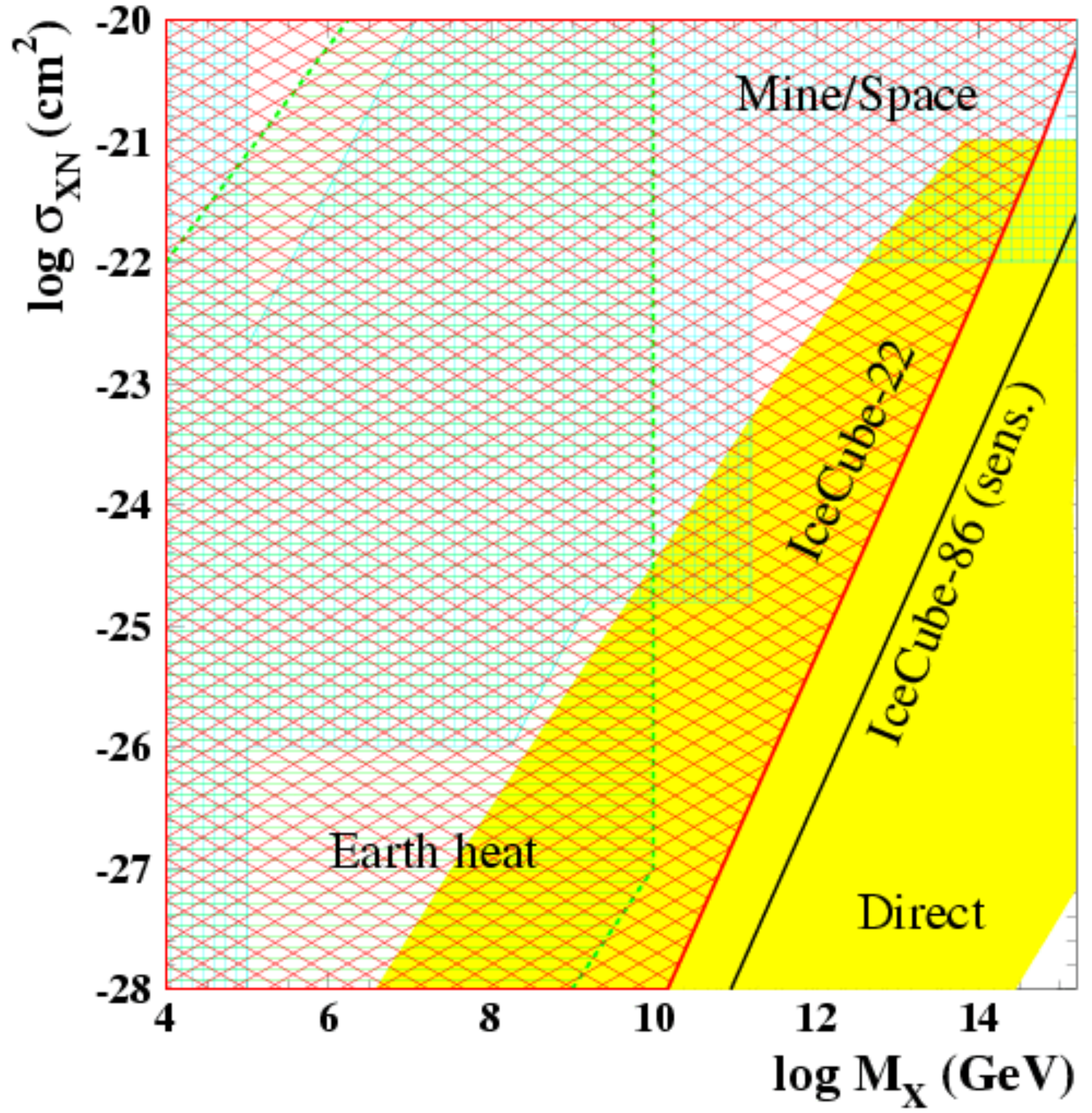}
\caption{Excluded region at 90\% C.L. in the simpzilla mass versus cross section
parameter space. The region labeled ``Direct'' (solid yellow region) was
excluded based on direct detection~\cite{Albuquerque:03a}; the ``Earth heat''
region (green striped region) is excluded based on the Earth's heat flow 
\cite{Mack:07a} and the blue squared hatched ``Mine/Space'' region is based on
many experiments~\cite{McGuire:01a,Starkman:90a}. Figure taken from~\cite{Albuquerque_Cph:10a}.}
\label{fig:Simp_limit}
\end{figure}
For the neutralino search~\cite{IC22_WIMP}, five neutralino masses were simulated, and two annihilation 
channels for each, a {\it hard} channel (annihilation into W$^+$W$^-$), and a {\it soft} channel, (annihilation into 
$b\overline{b}$).  Figure~\ref{fig:IC22_flux_limits} shows the limits on the flux of muons from neutrinos produced in neutralino 
annihilations at the center of the Sun as a function of neutralino mass. The green shaded area represents flux predictions from 
presently allowed combinations of the MSSM parameters. 
The grey shaded area represents the parameter space that would still be allowed if direct searches would improve their sensitivity 
 1000 times from the current limits. 

The limit on the muon flux can be converted into a limit on the spin-dependent neutralino-proton scattering cross section, 
$\sigma_{\chi+p}^{SD}$,  
since the muon flux is proportional to the capture rate of neutralinos in the Sun (mainly a proton system), and this quantity is in 
turn related to $\sigma_{\chi+p}^{SD}$. The conversion  can be uniquely performed by assuming equilibrium between 
capture and annihilation. A conservative limit is obtained by further assuming that 100\% of the cross section is 
spin-dependent~\cite{Wikstrom:09a}.  
The result of this exercise is shown in Figure~\ref{fig:IC22_Xsection_limits}, which shows the limits obtained with the 22-string 
IceCube detector on the spin-dependent neutralino-proton cross section as a function of WIMP mass. Current best limits from direct 
searches are also shown for comparison. The figure illustrates the capability of indirect searches to set strong limits on the 
spin-dependent cross section.\par 
 The same data set was used the the LKP search~\cite{IC22_LKP}. Seven LKP masses were simulated between 250~GeV 
and 3000~GeV and a similar analysis procedure as in the WIMP search was followed to obtain a limit on the LKP-proton spin-dependent 
cross section as a function of LKP mass, as shown in Figure~\ref{fig:LKP_limit}. The color code of the theoretical parameter space is 
as mentioned for Figure~\ref{fig:IC22_flux_limits}.

 The results of the searches presented above were interpreted in~\cite{Albuquerque_Cph:10a} in terms of superheavy dark matter. 
 In order to calculate the expected neutrino flux at the detector from Simpzilla annihilations in the Sun as a function 
of energy, the injection spectrum at the core of the Sun calculated in~\cite{Albuquerque_etal:01a} was used. 
The basic assumption is that Simpzillas annihilate into a pair of quarks or gluons, which will fragment into high multiplicity 
jets of hadrons. While most of the secondary products will lose energy in the dense solar interior before decaying into 
neutrinos, top quarks, due to their short lifetime, will decay into W$b$ without any energy losses, and the subsequent decay 
of the W bosons will produce a high energy neutrino flux in neutrino telescopes. The same data set and cuts used to 
extract limits on WIMPs can be used to set a limit on the Simpzilla-nucleon cross section. The results are shown in 
Figure~\ref{fig:Simp_limit}, which shows the exclusion region of the spin-dependent Simpzilla-nucleon cross section versus 
mass. The results of IceCube disfavour a big area of parameter space, complementary to direct searches and other previous 
results from detectors in space~\cite{McGuire:01a,Starkman:90a} or arguments based on heat generation at the center of the 
Earth~\cite{Mack:07a}. Taken all together, the results of these searches strongly disfavour super-heavy dark matter a solution 
to the galactic dark matter problem.

\begin{figure}[t]
\centering\includegraphics[width=0.9\linewidth,height=0.85\linewidth]{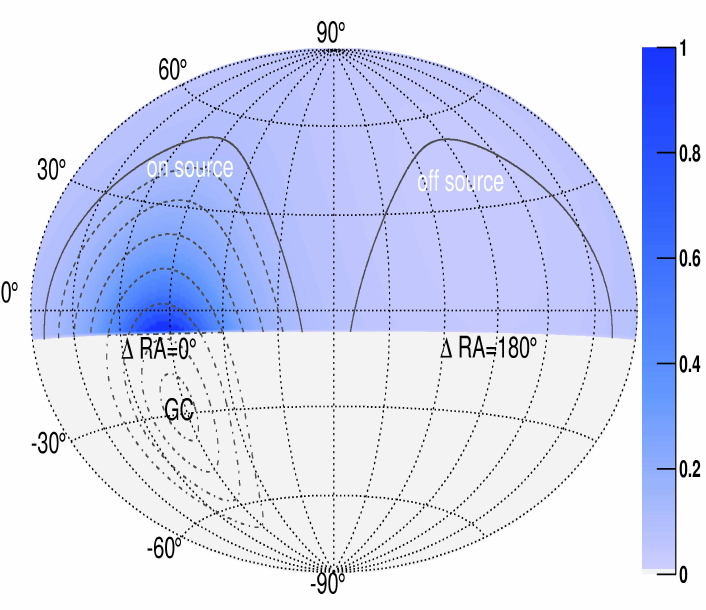}
\caption{Relative intensity of the expected neutrino flux from dark matter annihilations in the halo. The scale is 
normalized to 1 at the Galactic Center position, assuming a NFW density profile. 
The on-- and off--source regions for the Galactic Halo analysis are defined as shown by the solid lines. The on--source region 
is centered in RA around the Galactic Center and the off--source region is rotated by 180$^\circ$ in RA.}
\label{fig:skymap}
\end{figure}

\begin{figure}[t]
\centering\includegraphics[width=\linewidth,height=0.9\linewidth]{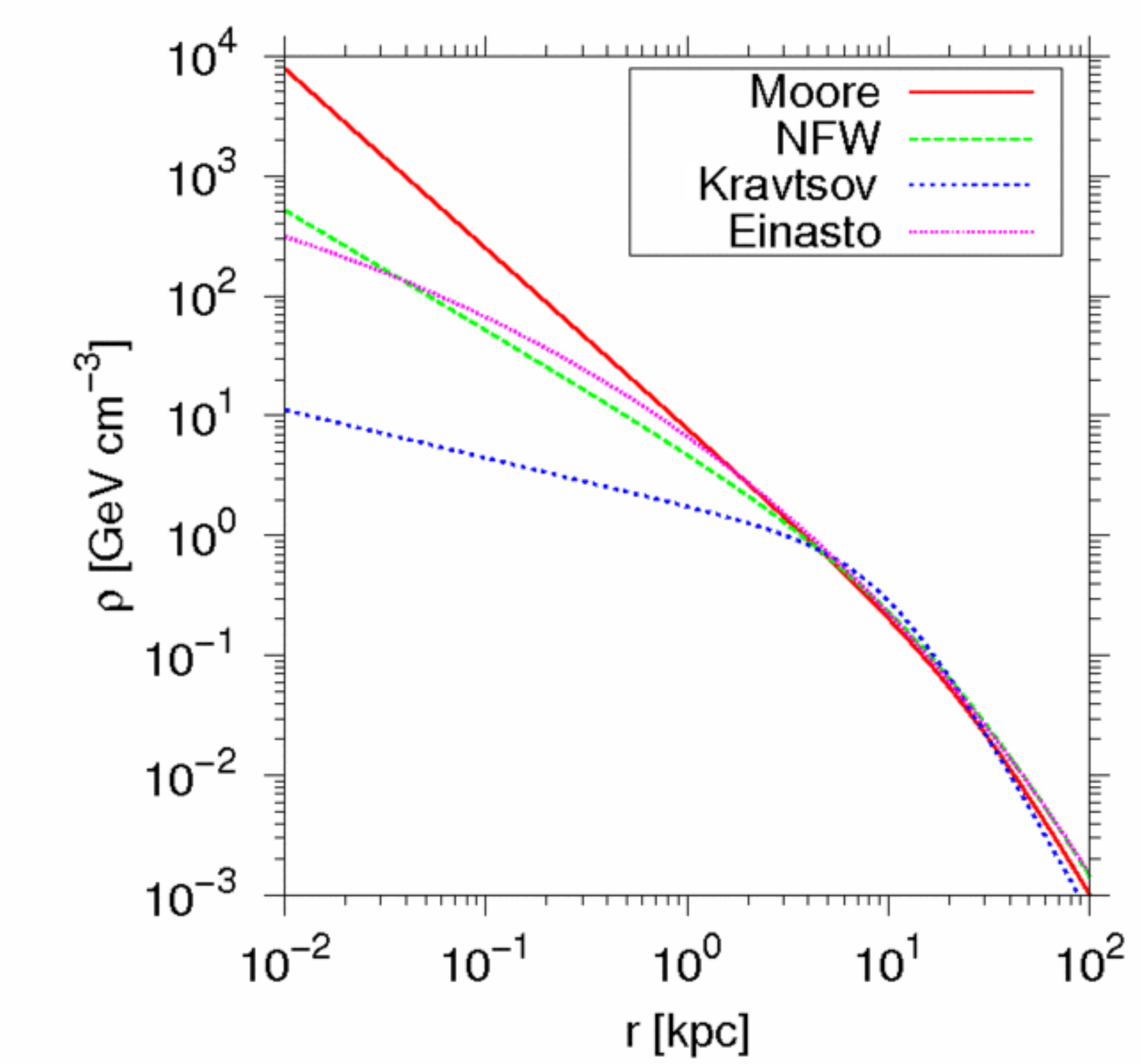}
\caption{Comparison of the predicted dark matter density distribution as a function of distance to the Galactic Center  
for the four models considered in the analyses presented in this note.}
\label{fig:halo_models}
\end{figure}
\section{Searches for dark matter signals from the Milky Way}
The Milky Way itself is a promising candidate to detect the products of dark matter annihilations in its center or halo. 
In this case neutrino telescopes are sensitive directly to the velocity-averaged dark matter self-annihilation cross section, 
$\left<\sigma_A v\right>$,  and the interpretation of the results is less dependent on issues related to capture of the dark matter 
particles through their interaction with normal matter. Also, the neutrino flux reaches the detector directly, without energy losses in 
the propagation through the dense solar interior, and a given annihilation channel produces a harder neutrino spectrum at the detector 
than in the solar case.\par

The expected neutrino flux from self-annihilations of a WIMP of mass m$_\chi$ is proportional 
to the square of the local dark matter density, $\rho^2$, integrated along the line of sight, 
J($\psi$), for a given angular distance from the Galactic Center $\psi$, and it is given by~\cite{Yuksel:07a}:
\begin{equation}
\frac{dN_\nu}{dE\,dA\,dT} = \frac{\left<\sigma_A v\right>}{2}J(\psi)\frac{R_\circ \rho^2}{4\pi m_{\chi}^2}\left(\frac{d N^\oplus_{\nu}}{dE}\right),
\label{eq:galaxyflux}
\end{equation}

where $dN^\oplus_{\nu}/dE$ is the neutrino energy 
distribution at the Earth, which depends on the supersymmetric model under consideration, and $R_{\circ}$ is the radius of the solar system 
orbit in the galaxy (8.5 kpc) and it normalizes the local dark matter content. We have used DARKSUSY~\cite{DARKSUSY} to 
calculate the neutrino flux from neutralino annihilations and we have chosen a few representative annihilation channels to 
derive $dN^\oplus_{\nu}/dE$: $b\bar{b}, W^+W^-, \mu^+\mu^-$ and $\nu\bar{\nu}$, assuming 100\% branching ratio to each channel in turn.
The expected number of neutrino events in the 
detector is then given by the integration of equation~\ref{eq:galaxyflux} over the detector live time and effective area A. 
This procedure allows the direct conversion between a limit on the expected flux at the detector and the self annihilation 
cross section. The prediction is in principle dependent on the chosen distribution of the dark matter through $R_{\circ}$ and 
$\rho^2$. There are different models in the literature to describe the distribution of dark matter in galaxies, based on 
N-body cold dark matter simulations or gravitational lensing observations~\cite{NFW,Einasto,Moore,Kravtsov}. These models 
generally show very similar behavior at large distances from the Galactic Center, but they differ significantly (up to orders 
of magnitude) in their predictions close to it, as shown in Figure~\ref{fig:halo_models}. 
In what follows, we use the NFW profile~\cite{NFW} as a benchmark for the analyses.

\subsection{Searches from the Galactic Halo}  
Even if the dark matter density in the galaxy is expected to peak towards the Galactic Center (which lies below the horizon at the location 
of IceCube), high statistics N-body simulations of galaxy formation with a dark matter component predict that the dark matter halo extends to 
several visible radii away from the center of the galaxy~\cite{Diemand:07a}. This opens the opportunity to use the Galactic Halo as an extended 
source of dark matter annihilations, where the signal would consist of a large-scale anisotropy in the IceCube neutrino 
sky~\cite{IC_Halo}. In order to perform such an analysis we have divided the northern hemisphere in two areas as depicted in
 Figure~\ref{fig:skymap}, the on-source region, containing the same RA as the Galactic Center, and the off-source region, 
rotated 180$^\circ$ in RA. An anisotropy in the neutrino flux would result in an excess of events 
$\Delta N = N_{\texttt{\tiny on}}-N_{\texttt{\tiny off}}$, where  $N_{\texttt{\tiny on}}$ 
and $N_{\texttt{\tiny off}}$ are the neutrino events counted in the on- and off-regions respectively.  
Systematic effects on the background estimation in the signal region are minimized since the on-- and off--source regions are of equal size. 
Remaining systematics are related to the uneven exposure as a function of azimuth angle, due to the uneven shape of the detector and gaps 
in the data sample after data cleaning. The effect is negligible though, being about 0.1\%. Another source of systematic 
uncertainties in the background estimation is the anisotropy of the cosmic ray angular distribution~\cite{IC_anisotropy}, but 
it is also a small (0.2\%) effect. \par
\begin{figure}[t]
\centering\includegraphics[width=0.95\linewidth,height=0.95\linewidth]{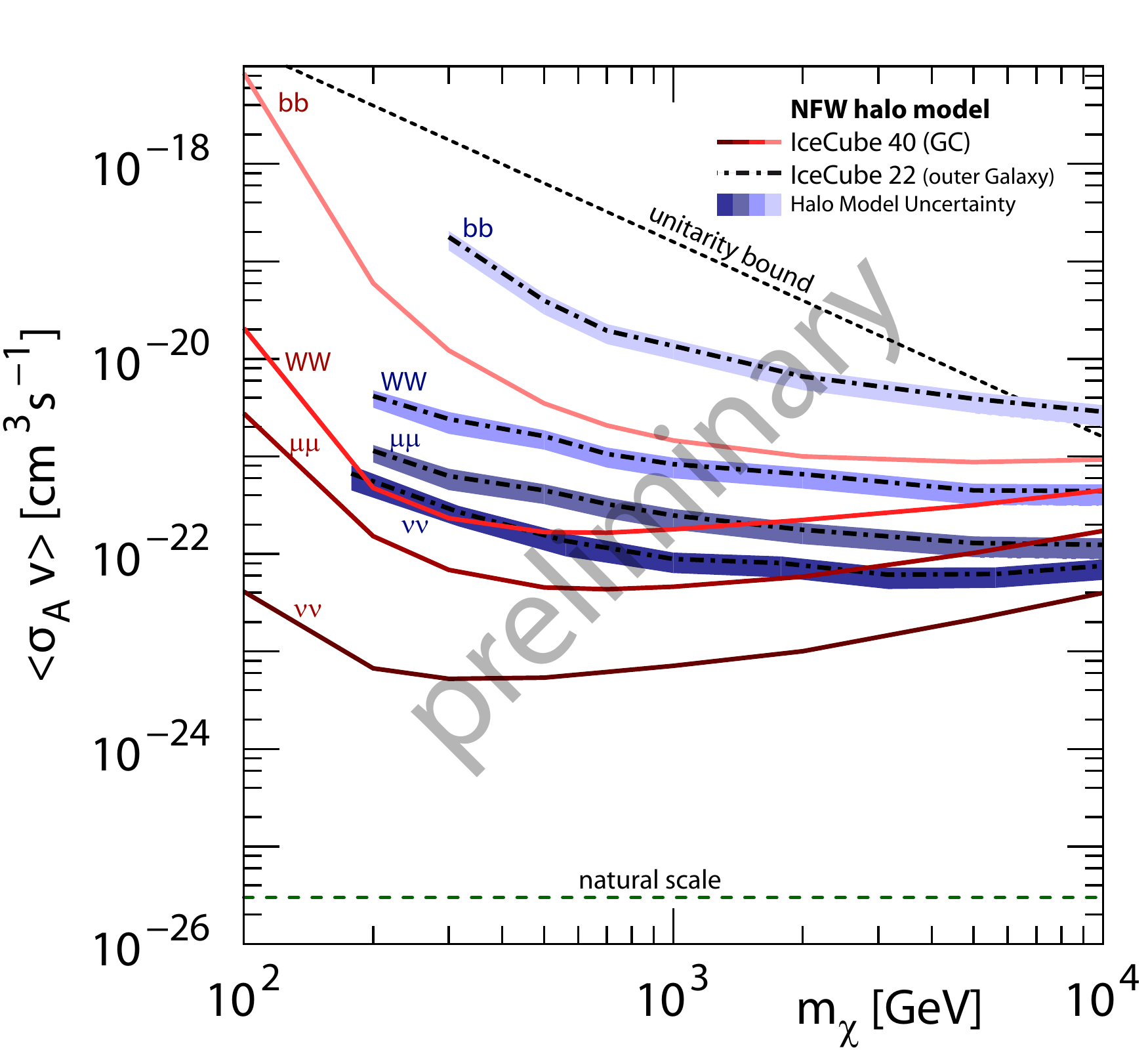}
\caption{90\% CL limits on the  $\left<\sigma_A v\right>$ versus neutralino mass from the 22-- and 40--string IceCube configurations. The 
40--string detector limits (thin lines) correspond to the Galactic Center analysis, while the 22-string detector limits (thick lines) correspond to the Galactic Halo analysis. 
The thickness of the halo analysis results represents the uncertainty due to the choice of the halo model. Central values (dot-dashed lines) are obtained with the NFW halo profile.}
\label{fig:halo_limits}
\end{figure}

 For this search we have used the sky map derived  in the IceCube 22-string point source analysis~\cite{IC_PSanalysis}.
The data set covers the whole northern sky with 5114 neutrino candidate events collected in 275.7 days of live time 
acquired during 2007-2008.  The data  contains 1367 events in our on-source region and 1389 in the off-source region, 
making $\Delta N$ compatible with the null hypothesis and allowing to set a limit on the neutrino flux from neutralino 
annihilations in the Galactic Halo, the left-hand side of equation~\ref{eq:galaxyflux}, and therefore a limit on $\left<\sigma_A v\right>$.  
Figure~\ref{fig:halo_limits} shows the limits obtained 
for the different annihilation channels studied, and it is discussed in more detail in the next section.\\

\begin{figure}[t]
\centering\includegraphics[width=\linewidth,height=0.85\linewidth]{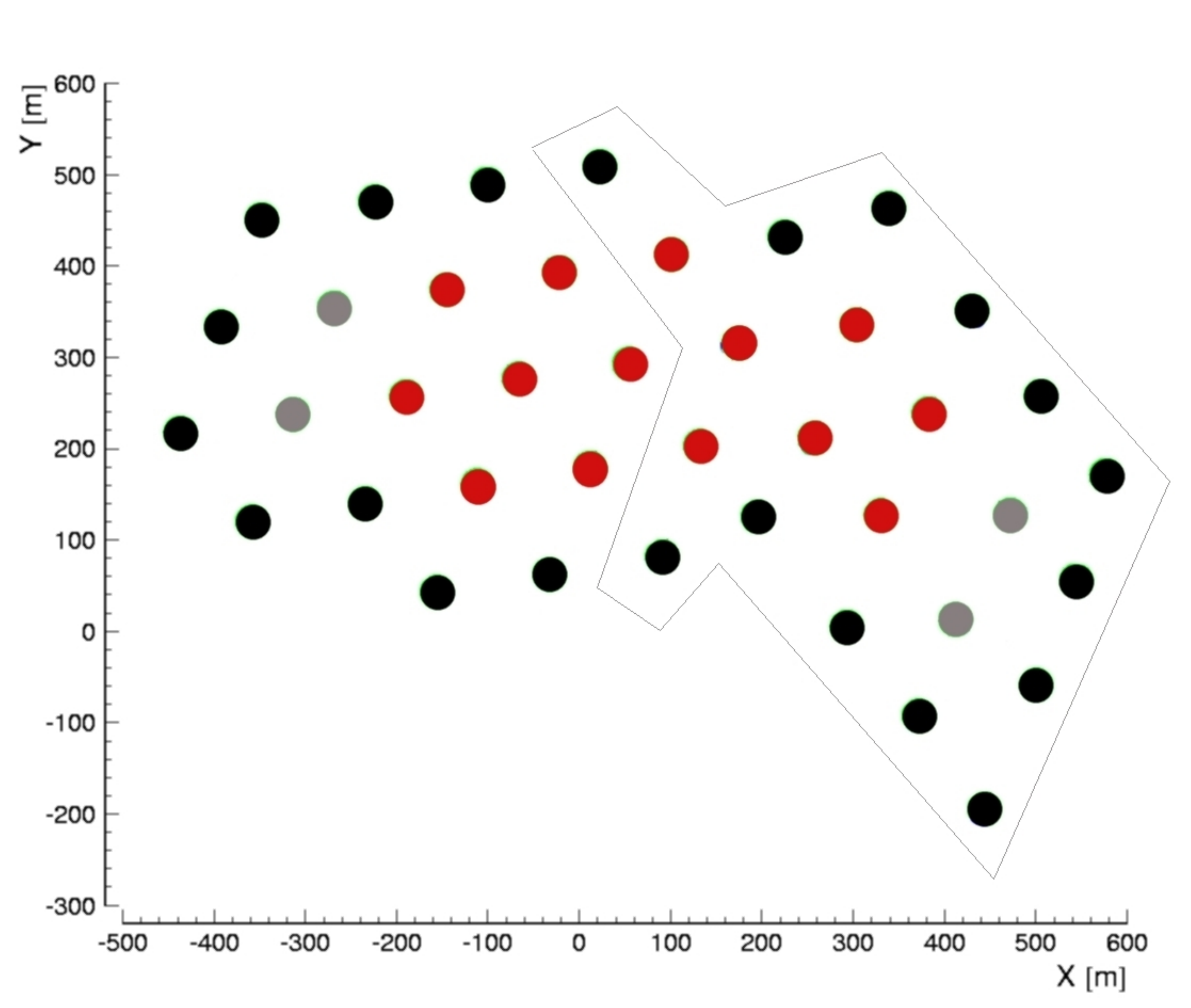}
\caption{The surface layout of the 40-string configuration of IceCube. For the Galactic Center dark matter analysis, the 
outer layer of strings (black dots) were considered as a veto, the first hit of the events required to be
 recorded in the inner core of strings (red dots). The  line delimits the 22-string configuration of IceCube, used in the solar and 
halo analyses.}
\label{fig:IC40geo}
\end{figure}

\begin{figure}[t]
\raisebox{1.8cm}{\centering\includegraphics[width=\linewidth,height=0.4\linewidth]{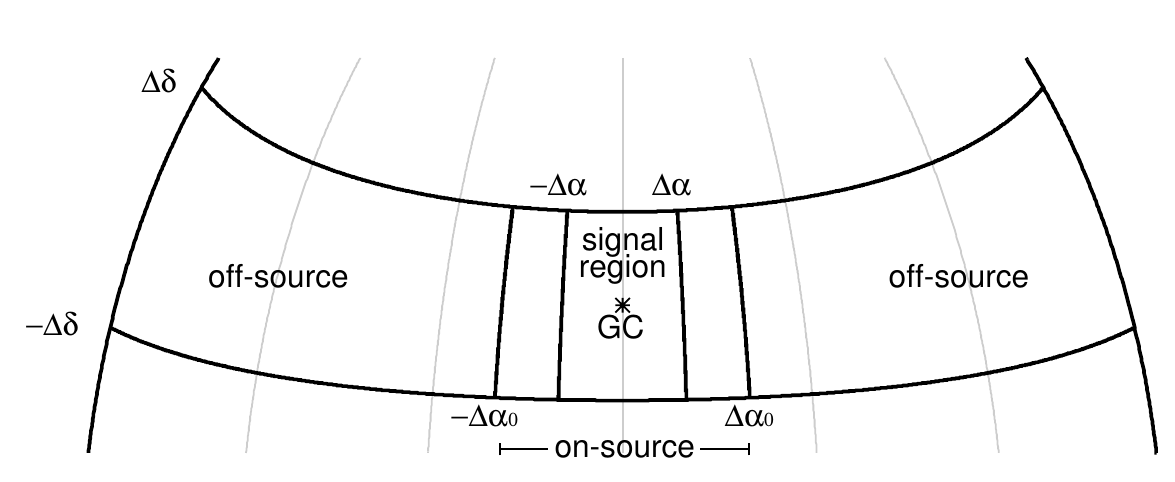}}

\caption{The definition of the on-- and off--source regions, as well as the signal region, used in the Galactic Center search (not to scale). 
The width of the off-source region in declination, $\Delta \delta$, is 8$^\circ$. $\Delta\alpha_o$ is taken as 20$^\circ$}
\label{fig:gc_on-off}
\end{figure}

\begin{figure}[t]\centering\includegraphics[width=\linewidth,height=\linewidth]{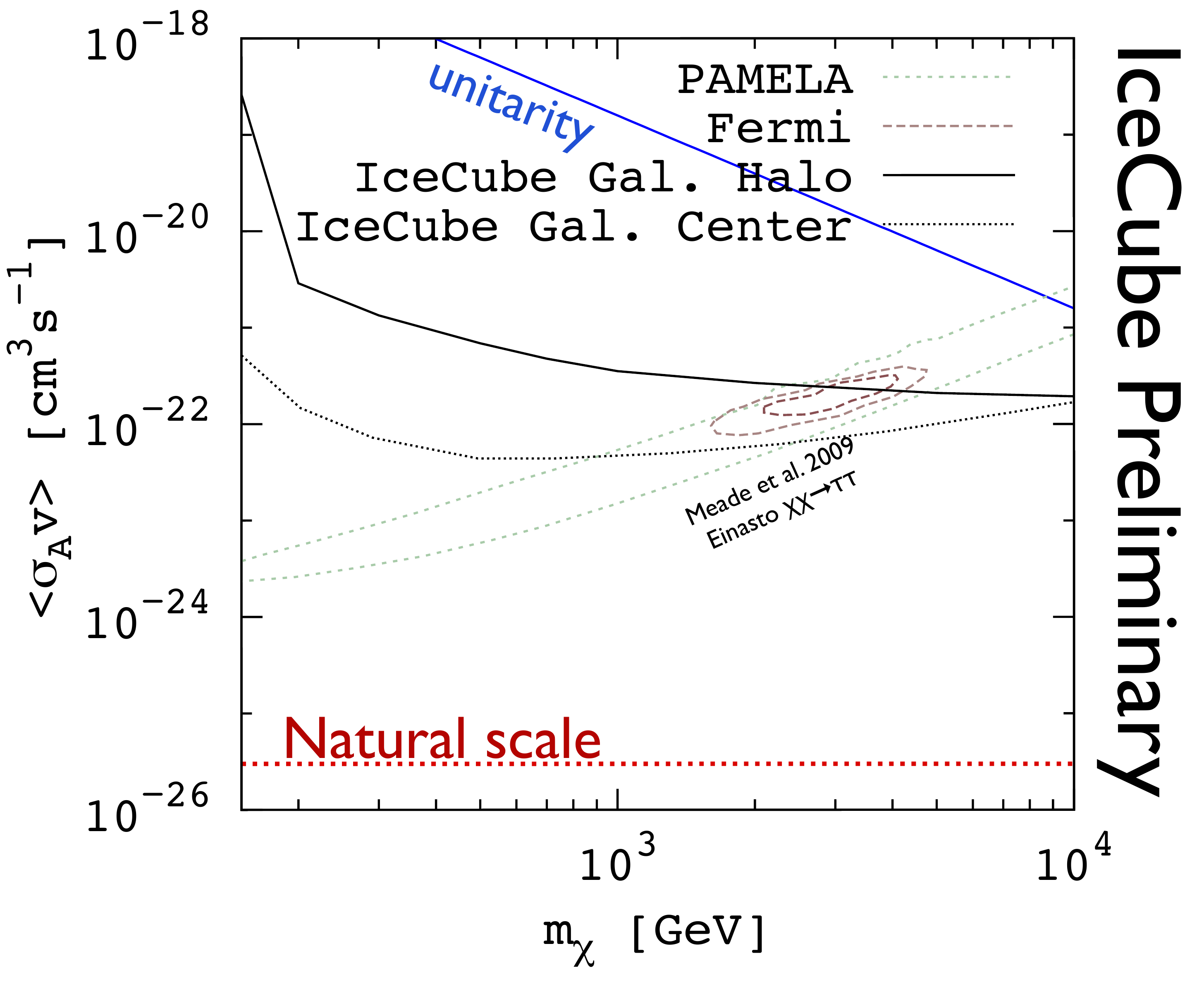}
\caption{The IceCube 90\%CL limits on $\left<\sigma_A v\right>$ assuming annihilation into $\tau^+\tau^-$, compared with the best fit region for the same 
model, using data from the Pamela and Fermi satellites (figure adapted from~\cite{Meade:09a}). The Einasto halo profile has been used in the derivation of 
results shown.}
\label{fig:comparison_FermiPamela}
\end{figure}

\subsection{Searches from the Galactic Center}
 The Galactic Center, (266$^\circ$ RA, -29$^\circ$ DEC), lies in the southern hemisphere viewed from the IceCube location, 
and any neutrino search from this region will be strongly affected by the presence of an overwhelming atmospheric muon background 
from above the horizon. 
 In order to reject this background we define a fiducial detector volume surrounded by a veto region, and consider only events with 
no hits on the veto region. This allows to define starting tracks which are the result of a neutrino interaction within the 
fiducial volume.  The 40-string IceCube detector was large enough so that defining an efficient veto geometry was possible 
for the first time. The veto consisted of the top thirty optical modules in each string, as well as the complete strings in the outer layer of 
the detector, shown as the black dots on Figure~\ref{fig:IC40geo}. 
The first hit of the event was required to happen in the inner core of the detector, marked by the red dots in Figure~\ref{fig:IC40geo}. 
An on-source/off-source search was then performed~\cite{Jan-Patrik:10a}. A search bin of $\Delta\delta=\Delta\alpha=\pm$8$^\circ$ in right ascension and declination 
around the Galactic Center was optimized as the signal region, and the rest of the declination band was taken as the off-source region for 
background determination. The optimal value of $\Delta\delta$ was shown not to depend strongly on the WIMP mass and annihilation channel 
considered, and the common value of 8$^\circ$ was chosen for all cases. 
A buffer zone of 12$^\circ$ from the edge of the signal region was left out of the analysis in order to avoid any possible 
signal contamination in the background region, see Figure~\ref{fig:gc_on-off}. 
The analysis was performed on data taken between April 2008 and May 2009, corresponding to a total of 367 d of detector live time. The 
technique used is the same as in the halo analysis, $\Delta N = N_{\texttt{\tiny on}}-N_{\texttt{\tiny off}}$ is calculated from the 
on-source and off-source regions and checked against the null hypothesis. The number of events obtained in the signal region 
was 798842, while 798819 events were expected from the off-source background estimation. In a similar way as in the halo 
analysis, a limit on   $\left<\sigma_A v\right>$ can be obtained from the non-observation of any event excess from the center 
of the Galaxy. \par

 Figure~\ref{fig:halo_limits} summarizes in one plot the preliminary limits on  $\left<\sigma_A v\right>$ obtained with both the Galactic Halo 
(thick blue-shaded lines) 
and the Galactic Center analyses (thin red lines). The thickness in the Galactic Halo results represents the small effect of the choice of the halo model in this analysis, 
the dashed lines being the result of the NFW choice. This is to be expected from Figure~\ref{fig:halo_models} where it can be seen that at the 
location of the Solar System (8.5 kpc) from the Galactic Center, all models predict a very similar dark matter density. 
The natural scale line marks the threshold under which the self annihilation cross section has values that 
are not interesting anymore to make the neutralino a good dark matter candidate. Systematic uncertainties in the signal and  background 
predictions are included in the limit calculation only for the Galactic Halo analysis. The effect of systematic uncertainties on the results of 
the Galactic Center analysis has still to be evaluated. \par
 
 Indirect dark matter searches with neutrinos from the Galactic Center and Halo allow a direct comparison with searches performed with 
photons or cosmic rays by detectors on space, opening the exciting possibility of a ``multi-wavelength'' dark matter search. The same 
annihilations that produce a neutrino flux are predicted to produce photons and electrons that can be measured by the Fermi and Pamela 
satellites. However, a photon or cosmic ray signal from dark matter annihilation in the Galactic Center region is plagued with uncertainties on the 
modeling of known background sources, which is not the case for the neutrino signal. 
 Indeed recent results by Fermi on the e$^+$e$^-$ and $\gamma$ flux~\cite{Fermi:09a} and by Pamela on a positron excess from the 
Galactic Halo~\cite{Pamela:09a} have been interpreted in terms of annihilating dark matter (see for example~\cite{Meade:09a} and references therein). 
Although alternative conventional scenarios exist to explain these results (in terms of pulsars or cosmic ray origin for example~\cite{Grasso:09a,Mertsch:09a}), 
a wealth of models based on the dark matter hypothesis have been proposed. Figure~\ref{fig:comparison_FermiPamela} shows the IceCube limits on 
$\left<\sigma_A v\right>$ as a function of neutralino mass, assuming the $\tau^+\tau^-$ annihilation channel. This is one of the channels 
that produce a hard neutrino spectrum from the decay of the $\tau$, and therefore a channel where high energy neutrino telescopes are competitive. 
The IceCube limits are overlayed with the best-fit region in the same phase space obtained from the Fermi and Pamela measurements in~\cite{Meade:09a}, and 
it can be seen that both the IceCube Galactic Halo and Galactic Center analyses reach the level of the best fit to Fermi and Pamela data. Specifically, 
the Galactic Center analysis  disfavours values of $\left<\sigma_A v\right>$ above about ~10$^{-22}$ cm$^3$s$^{-1}$, which precisely covers the 
 90\% CL contour of the fit to the satellite data.

\section{Conclusions}
 IceCube has an active program of searches for dark matter, both from candidates accumulated in the Sun as well as in the 
Galactic Halo or center. We have tested the data from the 22-string and 40-string configurations of IceCube for 
an excess neutrino flux from these objects and interpret the results in terms of several dark matter candidates. 
With the 40-string detector we have been able to search the Galactic Center for the first time . The size of the detector allows to use 
a fraction of the instrumented volume as veto region, which enables the identification of starting tracks and an efficient reduction of 
the atmospheric muon background. This technique 
will be used in its full potential with the complete 86-string detector in the future. The low-energy extension DeepCore 
which has been already deployed in the center of the IceCube array will allow to significantly lower the energy threshold of IceCube 
and extend the dark matter searches in a competitive way to the interesting region of candidate masses below 100~GeV.

\end{document}